\documentclass[usenatbib]{mn2e}
\usepackage{graphicx}
\begin{document}

\title[Benchmark experiments for exoplanetary GCMs]{Benchmark experiments with global climate models applicable to extra-solar gas giant planets in the shallow atmosphere approximation}
\author[V. L. Bending, S. R. Lewis, and U. Kolb]{V. L. Bending,$^1$ S. R. Lewis,$^1$ and U. Kolb$^1$ \\
$^1$ Department of Physical Sciences, The Open University, Milton Keynes MK7 6AA}

\maketitle
\begin{abstract}
The growing field of exoplanetary atmospheric modelling has seen little work on standardised benchmark tests for its models, limiting understanding of the dependence of results on specific models and conditions. With spatially resolved observations as yet difficult to obtain, such a test is invaluable. Although an intercomparison test for models of tidally locked gas giant planets has previously been suggested and carried out, the data provided were limited in terms of comparability. Here, the shallow \textsc{puma} model is subjected to such a test, and detailed statistics produced to facilitate comparison, with both time means and the associated standard deviations displayed, removing the time dependence and providing a measure of the variability. Model runs have been analysed to determine the variability between resolutions, and the effect of resolution on the energy spectra studied. Superrotation is a robust and reproducible feature at all resolutions.
\end{abstract}

\section{Introduction}\label{SecIntro}

Over the past several years, work has been carried out in the field of three-dimensional numerical simulation of exoplanetary atmospheres using large-scale global climate models (GCMs), as reviewed in, e.g., \citet*{ShowmanMenouChoReview}. In such modelling, it is important to establish clear model benchmarks, such as that set up for Earth by \citet{HeldSuarez}. Little benchmark work has been carried out for exoplanetary studies, however, with the exception of that initially described by \citet{MenRauIntercomp}, who laid out an intercomparison test and investigated the response of the spectral \textsc{igcm} dynamical core to it, and further studied by \citet*{intercomp2}, who investigated the response of both spectral and gridpoint cores available for implementation in the \textsc{fms} model to both this and additional tests. This study follows the lead provided by these two papers in carrying out a clear and reproducible intercomparison test, and adds to it further statistics and information, including online data, to facilitate model comparisons.

Although many modelling studies have been carried out on the atmospheres of `hot Jupiters', gas giant planets less than about 0.1 AU from their host star \citep[e.g.][]{IYAexoreview}, few direct comparisons of model responses to these conditions have been made. It is imperative that model-dependent responses be identified and understood. Without such intercomparison studies, it cannot be determined which elements of a simulation may correspond to the planet under study, and which are simply artefacts of a specific model, or a result of the high sensitivity of these complex, non-linear systems to the precise input parameters and initial conditions. \citet{TCInitFlow} have studied the effects of initial flow on the final state reached by their model, at atmospheric depths from 1 bar down to 100 bar, and found the final state to be highly dependent on the initial conditions chosen. In contrast, \citet{LSInit} carried out a similar study down to 200 bar, and found almost no dependence on the initial conditions, a result that \citet{LSInit} suggest may be due in part to the different vertical profiles of the thermal forcing and the restoration timescale between the two studies. With little observational information available, it becomes vitally important to have a reference simulation against which all models can be compared. Whereas models of Solar System planets may be compared to observations of the planet in question to determine their effectiveness, the modelling of extra-solar planets must proceed from first principles, though in some cases phase curves may be utilised to gain information on the conditions at the photosphere.

A wide variety of models are utilised in simulating hot Jupiter atmospheres, from `shallow-water' models such as that used in \citet{ShowmanSuperrot} to three-dimensional GCMs of varying complexity and underlying assumptions. Atmospheric depths studied range from 1 bar (`shallow') to 100s of bars (`deep'), and model atmospheric equations range from the one-layer shallow-water equations through the shallow, hydrostatic primitive equations to the full three-dimensional Navier-Stokes equations \citep[e.g.][]{IDD-3dNS}. The \textsc{sparc/mit}gcm of \citet{ShowmanSPARC} couples a correlated-\textit{k} radiative transfer implementation to their dynamical core, while other models may utilise dual-band `grey' radiative transfer \citep*[e.g.][]{HFPrad}, or omit radiative transfer and utilise instead Newtonian relaxation towards a predetermined temperature state.

Large variability between models shows that some aspects of the situation have not as yet been positively determined using such models in their present form. Since the input parameters of even simple GCMs are poorly constrained by the available data on the planets to which they are applied, the choice of parameter adds a further degree of uncertainty to the already variable results. This again illustrates the importance of a single, fully-specified test case from which differing model-dependent responses may be determined.

This paper builds on work suggested by \citet{MenRauIntercomp} and further studied in \citet{intercomp2}, and adds the \textsc{puma} (Portable University Model of the Atmosphere) model \citep{PUMAref05} to those to which the test has been applied. Since the precise state of the model atmosphere is highly time-dependent, long-term statistics are produced to allow a quantitative comparison for future tests. Time means of the output fields are produced to gain an understanding of the overall characteristics of the modelled atmosphere, and the associated standard deviations to gain a quantitative understanding of the degree of variability in each field. These statistical analyses of the model output fields will allow future work to be compared on a more solid footing.

Section \ref{SecModelAndSetup} outlines the model used and the description of the experiment, including a full list of model parameters.
Section \ref{SecResults} displays the results of carrying out the intercomparison test with \textsc{puma}, and provides time mean and variability statistics. The data files required to produce these plots are provided online.
Section \ref{SecDiscuss} provides a discussion of these results and comparison to the previous results of \citet{MenRauIntercomp} and \citet{intercomp2}.
Section \ref{SecConclusions} covers the conclusions drawn, recommending the use of time mean and standard deviation statistics and a minimum sampling frequency, noting the degree of correspondence to the results of the previous authors, and suggesting additional plots to capture further aspects of the simulation not seen when studying only wind and temperature fields. In common with previous studies, and as expected from the work of \citet{ShowmanSuperrot}, a strong equatorial superrotation is found, resulting in an offset temperature hotspot.

\section{A model for gas giant exoplanets}\label{SecModelAndSetup}

\textsc{puma} is a simple global circulation model of the atmosphere, developed in its current form at the University of Hamburg \citep{PUMAref05} and descended from the spectral code of \citet{HoskinsSimmons75}. Though designed for use on Earth, the basic equations of the model are applicable to any system about which similar assumptions can be made, modelling an atmosphere in hydrostatic equilibrium that can be assumed to be a `thin shell' with respect to the radius of the (spherical) planet. These assumptions, notably the hydrostatic balance and shallow atmosphere approximations, result in a set of equations typically referred to as the primitive equations of meteorology. Its long heritage means that the model's dynamical core is well known and has been extensively tested, rendering it particularly useful for benchmark tests.

The model runs in spectral space in the horizontal using a triangular truncation, resolution being specified by the truncation coefficient, such as T21, which equates to a grid of approximately 64 longitudes and 32 latitudes. In general, if the truncation number is $T_\rmn{n}$, the number of latitudes is given by $N_\rmn{lat} = (3T_\rmn{n}+1)/2$. Particular resolutions, such as T42 ($128 \times 64$), T63 ($192 \times 96$), or T85 ($256 \times 128$), are widely used as standard resolutions owing to the resulting grid sizes being entirely or almost entirely powers of two, which allows for greater efficiency in the fast Fourier transform routines.

A finite-difference method is used in the vertical, with vertical levels defined using the sigma-coordinate \citep{SigmaRef}:
\begin{equation}
	\sigma = \frac{p(\lambda,\phi,z,t)}{p_\rmn{s}(\lambda,\phi,t)}
\label{EqSigmaExplanation}
\end{equation}
where $p$ is the pressure at any point in the atmosphere, and $p_\rmn{s}$ is the `surface' pressure at the same $\lambda, \phi, t$, with $\lambda$, $\phi$, $z$, and $t$ being longitude, latitude, height above the model base, and time, respectively. $\sigma$ is then always 1 at the base of the model atmosphere, decreasing upwards. Model levels $\sigma = l_\rmn{n}$, on which the variables are calculated, take values in the range $0 < l_\rmn{n} < 1$, with boundary conditions imposed at $\sigma = 0$ and $\sigma = 1$. In this gas giant case, the `surface' is flat, determined by a reference geopotential, and the boundary conditions equate to a rigid surface at the top and bottom of the atmosphere, at which the vertical velocity is set to zero. Any level distribution may be input; however, for the purposes of this experiment, only linear spacing in sigma was used.

Choosing a horizontal and vertical resolution requires a compromise to be made between the degrees of freedom of the model and the computing power required to run it within a feasible time-scale. Too low a resolution may result in misleading results as features vital to the flow are not represented. The resolution dependence of this experiment is discussed further in Section \ref{SecDiscuss}. The particular resolutions used in this paper are chosen to correspond with those of \citet{MenRauIntercomp} and \citet{intercomp2}.

\textsc{puma} has been modified for use with gas giant planets, permitting all planetary parameters to be changed, and also to permit the usage of three-dimensional, customised temperature restoration fields. Equations \ref{EqForcingFull} to \ref{EqForcingTtheta} were added as a selectable alternate option to the standard, Earth-like temperature restoration field. These modifications permit a tidally locked forcing scenario to be implemented, as is necessary for hot Jupiter type planets, with extreme temperature differences between the star-facing dayside and shielded nightside. \citep{MenRauIntercomp}.

\begin{equation}
T_\rmn{eq}( \lambda , \phi , \sigma ) = T^\rmn{vert}_\rmn{eq}(\sigma) + {\beta}_\rmn{trop} (\sigma) \Delta T_{\theta}( \lambda , \phi )
\label{EqForcingFull}
\end{equation}
where
\begin{eqnarray}
T^\rmn{vert}_\rmn{eq}(z) & = & T_\rmn{surf} - \Gamma _\rmn{trop} \left( z_\rmn{stra} + \frac{z - z_\rmn{stra}}{2} \right) \nonumber \\
 & & + \sqrt{ \left( \frac{1}{2} \Gamma _\rmn{trop} [z-z_\rmn{stra}] \right)^2 + {\delta T_\rmn{stra}}^2}
\label{EqForcingTvert}
\end{eqnarray}
\begin{equation}
{\beta}_\rmn{trop} (\sigma) = \left\{
\begin{array}{ll}
	\sin \left( \frac{\pi}{2}(\sigma - \sigma _\rmn{stra}) / (1 - \sigma _\rmn{stra}) \right) &
	\sigma \geq \sigma _\rmn{stra} \\
	0 & \sigma < \sigma _\rmn{stra} \\
\end{array}
\right.
\label{EqForcingBeta}
\end{equation}
and
\begin{equation}
\Delta T _{\theta} ( \lambda , \phi ) = \cos (\lambda) \cos (\phi) \Delta T_\rmn{EP}
\label{EqForcingTtheta}
\end{equation}

\begin{figure}
\includegraphics[height=\columnwidth,angle=90]{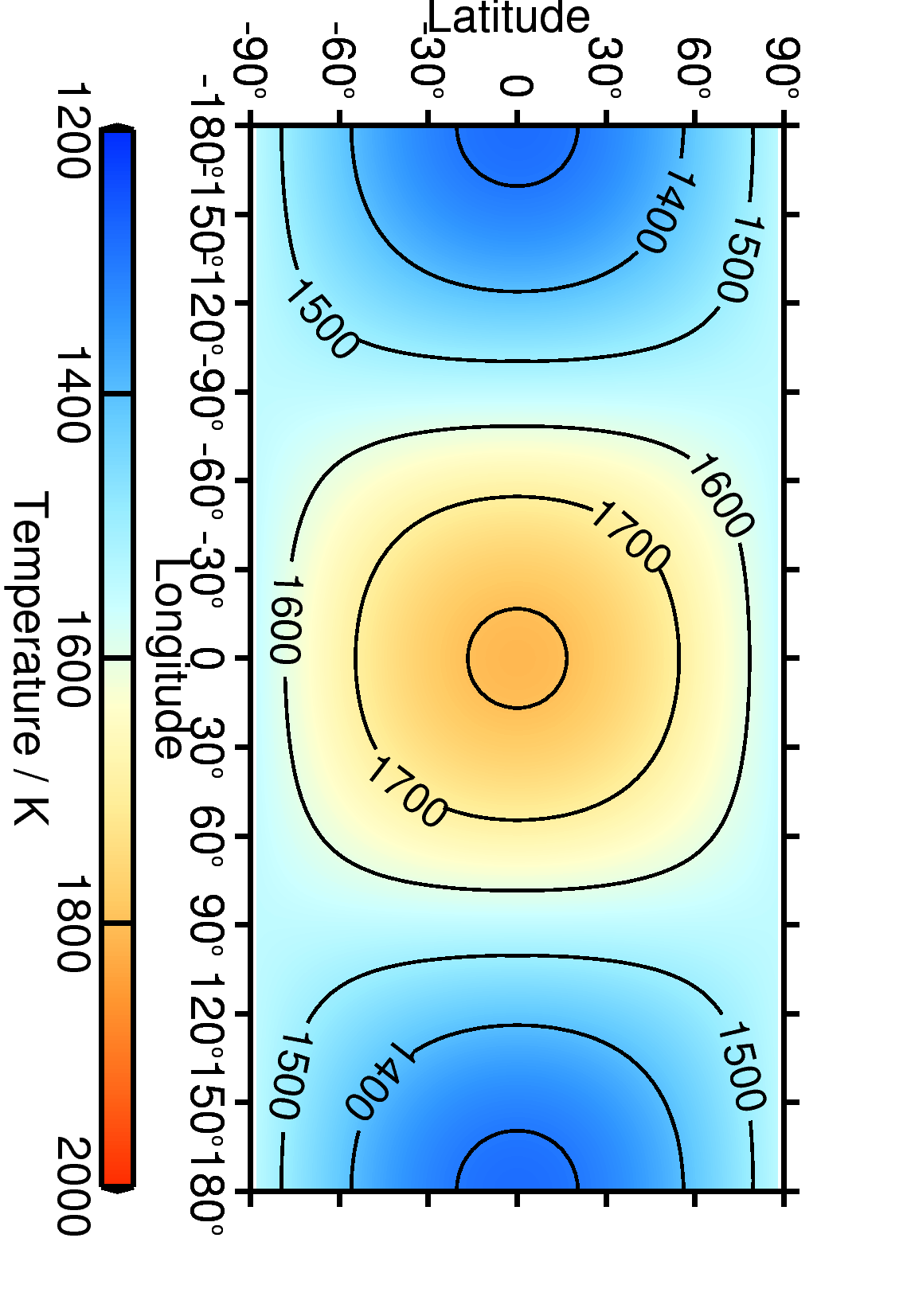}%
\caption{Latitude-longitude plot of forcing temperature for the T42L15 run at $\sigma = 0.7$, showing the dayside `hotspot' and cool nightside. The substellar point is at the centre of the image, and plots for other resolutions are identical.}
\label{FigT42LatLonForcing}
\end{figure}

\begin{figure}
\includegraphics[height=\columnwidth,angle=90]{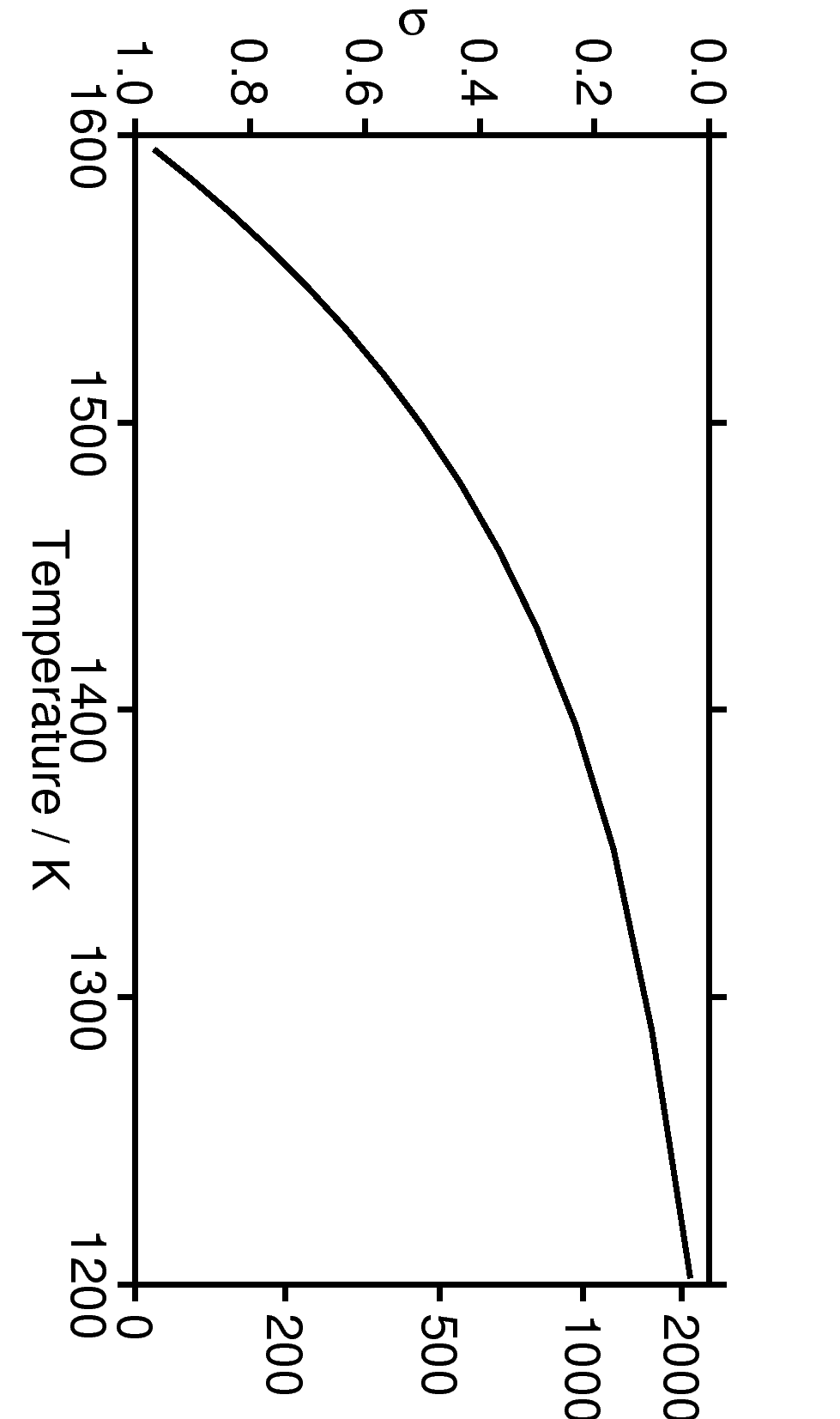}
\caption{Zonal mean plot of forcing temperature. The longitude averaging causes the day-night temperature difference to cancel out, so that the average equatorial temperature is equal to that at the poles.}
\label{FigT42ZMForcing}
\end{figure}

The temperature produced by radiative-convective equilibrium, without winds or other factors, is represented by $T_\rmn{eq}$, with the purely vertical part of this structure represented by $T^\rmn{vert}_\rmn{eq}$ and the purely horizontal part by $\Delta T_\theta$. The vertical element of the profile was chosen by \citet{MenRauIntercomp} to match that calculated by \citet*{IroRad} for HD 209458b. Such radiative-convective equilibrium temperatures may also be computed analytically from first principles using models such as that of \citet{GuillotRad}, which work has been extended to include the effects of scattering by \citet{HengRad}. A scaling factor $\beta_\rmn{trop}$ is applied to steadily decrease the temperature difference between the dayside and nightside until it becomes zero above the tropopause, represented by $\sigma_\rmn{stra}$ and $z_\rmn{stra}$. Similarly, the temperature increment at the tropopause is given by $\delta T_\rmn{stra}$ \citep{MenRauIntercomp}. The dry adiabatic lapse rate is represented by $\Gamma_\rmn{trop}$, and the mean `surface' temperature (temperature at the base of the model atmosphere) by $T_\rmn{surf}$, with the equator-to-pole temperature difference denoted by $\Delta T_\rmn{EP}$. Examples of the resulting temperature forcing pattern can be seen in Fig. \ref{FigT42LatLonForcing} and Fig. \ref{FigT42ZMForcing}. A `cold spot' is produced on the nightside of equal magnitude to the dayside hotspot. As a result, the zonal mean, or longitudinally averaged, forcing temperature is the same from equator to pole on every level of the atmosphere, producing the vertical structure shown in Fig. \ref{FigT42ZMForcing}.

Runs were carried out at T42, T63, and T85 resolutions, with 15, 20, and 20 levels respectively, linearly spaced in $\sigma$. A full list of model, planetary, and run-specific parameters can be found in Tables \ref{ModelTable}, \ref{PlanetTable}, and \ref{RunTable}, respectively. All parameters are taken from \citet{MenRauIntercomp}, who chose the planetary parameters to correspond to parameters for HD209458b, a well-studied inflated hot Jupiter, and \citet{intercomp2}.

\begin{table}
\caption{Table of model parameters}
\label{ModelTable}
\begin{tabular}{lcc}
  \hline
  Parameter & Symbol & Value \\
  \hline
  Ndel                               &                & 8                  \\
  Dissipation time-scale / day       & $t_\rmn{diss}$ & $5 \times 10^{-3}$ \\
	Rayleigh friction time-scale / day & $t_\rmn{frc}$  & $\infty$           \\
	Newtonian relaxation time /day     & $t_\rmn{rest}$ & $0.5$              \\
  \hline
\end{tabular}
\end{table}

The hyper-diffusion time-scale on the smallest resolved scale is given by $t_\rmn{diss}$, while Ndel gives the order of that diffusion. As a result of the conservation of energy built into these models, energy `builds up' at the resolution limit, requiring artificial diffusion to be applied to damp down and remove this effect, which would otherwise introduce unwanted effects into the results. It is important to note that the hyper-diffusion time-scale applies to the smallest resolved scale in each run, and thus results in a weaker dissipation on any given scale for a higher-resolution run as compared with a lower-resolution one, effectively producing weaker overall diffusion. The effects of this can clearly be seen in Fig. \ref{FigEnSpec}, where increasing resolution permits the model to represent an energy cascade to smaller and smaller scales before being cut off by diffusion. The order and magnitude of the hyper-diffusion values are non-physical to an extent and require tuning to the simulation being run. A more detailed discussion of the effects of forcing and dissipation, with particular relevance to simulations such as that carried out here, can be found in \citet{ChoForcDiss}.

Rayleigh friction applies an effective frictional force to drag the winds towards zero velocity on each level on which it is set, using the time-scale $t_\rmn{frc}$; a time-scale of infinity thus indicates that no friction is applied. The Newtonian relaxation time, $t_\rmn{rest}$, determines the rate at which the model temperature is forced towards the input radiative-convective equilibrium state $T_\rmn{eq}$.

\begin{table}
\caption{Table of planetary parameters}
\label{PlanetTable}
\begin{tabular}{lcr}
	\hline
	Parameter & Symbol & Value \\
	\hline
	Planetary radius / m                         & $a$                   & $10^8$    \\
	Rotation rate / $10^{-5}$ rad s$^{-1}$       & $\Omega$              & $2.1$     \\
	Gravity / m s$^{-2}$                         & $g_\rmn{p}$           & $8.0$     \\
	`Surface' pressure / bar                     & $P_\rmn{0}$           & $1.0$     \\
	`Surface' temperature / K                    & $T_\rmn{surf}$        & $1600$    \\
	Equator-pole temperature difference / K      & $\Delta T_\rmn{EP}$   & $300$     \\
	Tropopause temperature increment / K         & $\delta T_\rmn{stra}$ & $10$      \\
	Tropopause height / $10^6$ m                 & $z_\rmn{stra}$        & $2$       \\
	Adiabatic lapse rate / $10^{-4}$ K m$^{-1}$  & $\Gamma_\rmn{trop}$   & $2$       \\
	Specific gas constant / J kg$^{-1}$ K$^{-1}$ & $R_\rmn{s}$           & $3779$    \\
	Heat capacity / J kg$^{-1}$ K$^{-1}$         & $c_\rmn{p}$           & $13226.5$ \\
	\hline
\end{tabular}
\end{table}

The `surface', or base of the model, is defined as being at the 1-bar pressure, $P_\rmn{0}$. The gravitational constant throughout the model is given by $g_\rmn{p}$, and the specific gas constant and heat capacity at constant pressure by $R_\rmn{s}$ and $c_\rmn{p}$ respectively.

\begin{table}
\caption{Table of run-specific parameters}
\label{RunTable}
\begin{tabular}{lcccc}
	\hline
	Parameter & Symbol & T42L15 & T63L20 & T85L20 \\
	\hline
	No. of latitudes       & $N_\rmn{lat}$ & 64  & 96  & 128 \\
	No. of longitudes      & $N_\rmn{lon}$ & 128 & 192 & 256 \\
	No. of vertical levels & $N_\rmn{lev}$ & 15  & 20  & 20  \\
	\hline
\end{tabular}
\end{table}

Each run was carried out for a minimum of 350 planetary sidereal days following the determined spinup period, or approximately 1,000 Earth days \citep{intercomp2}. Records were made ten times per day, equally spaced from one another, and all time averaging periods are over the 350 days unless otherwise stated. In this paper, the term `day' shall always mean planetary sidereal days for consistency, noting that \citet{MenRauIntercomp} and \citet{intercomp2} differ in their use of the term `day', the former to mean planetary sidereal, the latter to mean Earth solar.

\section{Results}\label{SecResults}

\begin{figure}
\includegraphics[height=\columnwidth,angle=90]{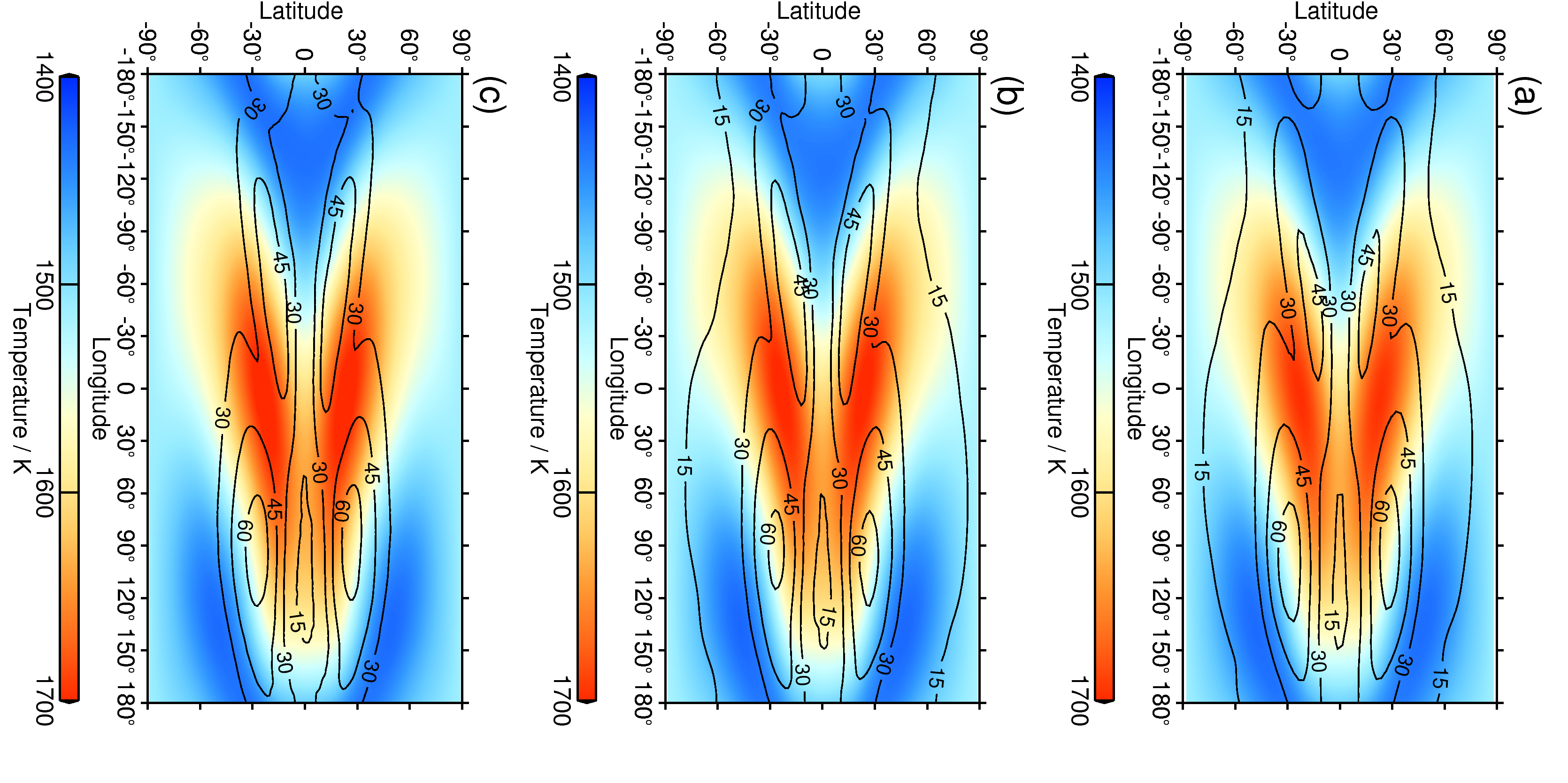}
\caption{Temperature mean and standard deviation for different resolution runs. Colours denote mean temperature, contour lines the standard deviation in Kelvin. The substellar point is at the centre of the image, or (0,0). (a) T42L15 at $\sigma = 0.7$, (b) T63L20 at $\sigma = 0.725$, (c) T85L20 at $\sigma = 0.725$.}
\label{FigLLMeanAndSD}
\end{figure}

Fig. \ref{FigLLMeanAndSD} shows the time mean temperature field in colour, with standard deviation contours overlaid. The model level closest to $\sigma = 0.7$ was chosen in each case, and all time means are taken over the 350-day period covering model days 30-380. It can be seen that the magnitude of both mean temperature and standard deviation is approximately consistent between runs, with standard deviation increasing slightly with resolution, in particular noting that the T85L20 run in plot (c) shows no 15 K standard deviation contour towards the poles. The differences in the shape of the standard deviation contours between T42L15 and T63L20 may also be partially due to the different locations of the model levels, at $\sigma = 0.7$ and $\sigma = 0.725$ respectively, an unavoidable result of using linear sigma spacing with different numbers of levels. The highest variance is found at $\pm 30^\circ$ N and approximately $100^\circ$ E of the substellar point, where the equatorial jet has carried warm air from the dayside to the cool nightside, making the fluctuations caused by its north-south movement most apparent. Though small differences are apparent, all runs demonstrate very similar mean temperature fields and standard deviations. The maximum mean temperature is 1698 K at T42, 1705 K at T63, and 1709 K at T85, with corresponding maximum standard deviations of 67 K at T42, 64 K at T63, and 66 K at T85.

\begin{figure}
\includegraphics[height=\columnwidth,angle=90]{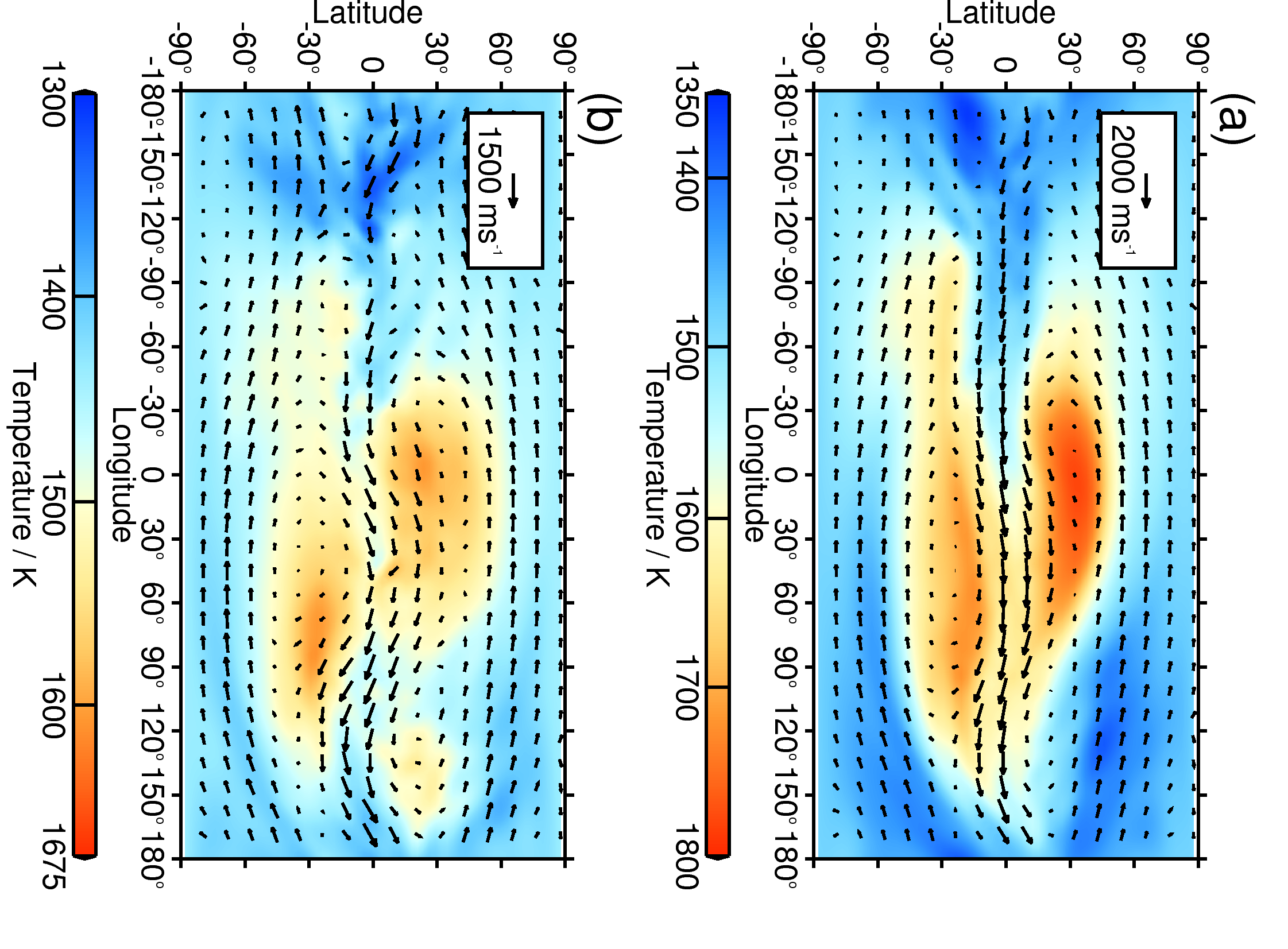}
\caption{`Snapshot' of T42L15 temperature and wind vectors for comparison with \citet{MenRauIntercomp}. The substellar point is at the centre of the image. (a) $\sigma = 0.7$, (b) $\sigma = 0.37$.}
\label{FigMRIEcylind}
\end{figure}

Versions of the T42L15 latitude-longitude instantaneous `snapshot' plots in fig. 3 of \citet{MenRauIntercomp} are also shown for comparison in Fig. \ref{FigMRIEcylind}. While they do not precisely replicate the pattern of those images, the same general features can be seen, with the strong equatorial jet and weaker reverse flow beyond $\pm 30^\circ$ N producing the temperature distribution seen in Fig. \ref{FigMRIEcylind}. While the winds appear similar, the lack of a scale in \citet{MenRauIntercomp} means that their strength cannot be directly compared in this image. Instead, reference must be taken from Fig. \ref{FigZWsnap}, which shows very similar zonal wind speeds to fig. 4 of \citet{MenRauIntercomp}. Although the mean state is robust, the precise form of the equatorial jet and the temperature distribution at a given moment is unpredictable, as the model is highly non-linear and thus its evolution is extremely sensitive, rendering such plots of limited use for the comparisons required in a benchmark test.

\begin{figure}
\includegraphics[height=\columnwidth,angle=90]{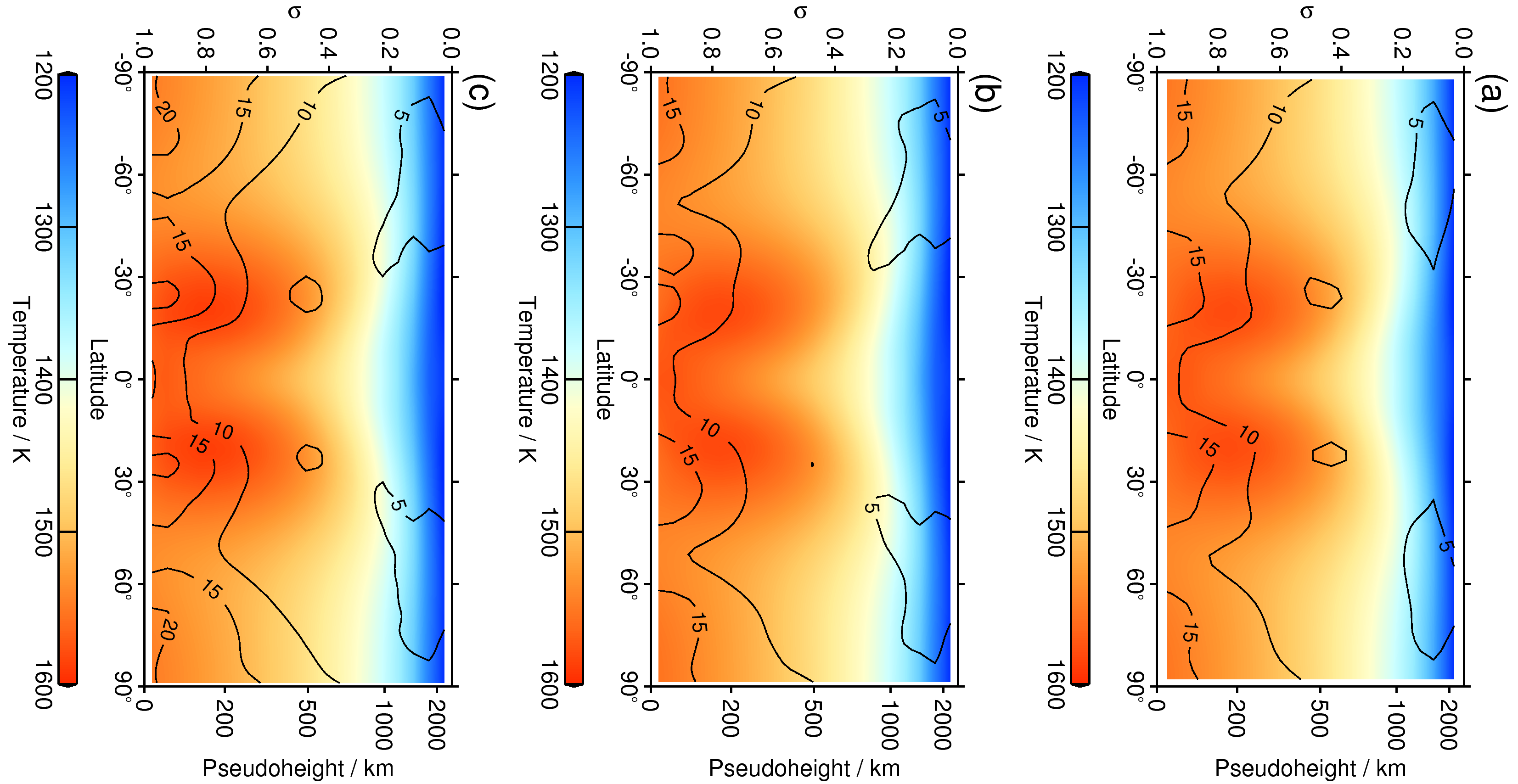}
\caption{Time-averaged zonal mean temperature plots at different resolutions. Colours denote mean temperature, contour lines the standard deviation in Kelvin. (a) T42L15, (b) T63L20, (c) T85L20.}
\label{FigZMTemps}
\end{figure}

The zonal mean temperature field, also averaged over the 350-day time period, is shown in Fig. \ref{FigZMTemps}. Note the different temperature scale to that of Fig. \ref{FigLLMeanAndSD}, due to the different temperature range encompassed. The most notable departure from the forcing state's simple vertical profile can be seen in the centre of the figure, between $\pm 40^\circ$ N, where it can be seen that two `hotspots' have developed, corresponding to the location of the temperature peaks seen either side of the equator in Fig. \ref{FigLLMeanAndSD}. These plots show the greatest difference in standard deviation, with the location of the contours differing noticeably, particularly towards the top and bottom of the model atmosphere, although this may to some extent be due to the limited resolution in the vertical. All runs are noticeably cooler towards the poles below around $\sigma = 0.8$ than those in fig. 6 of \citet{intercomp2}, which show temperatures of almost equal magnitude to those at the equator.

\begin{figure}
\includegraphics[height=\columnwidth,angle=90]{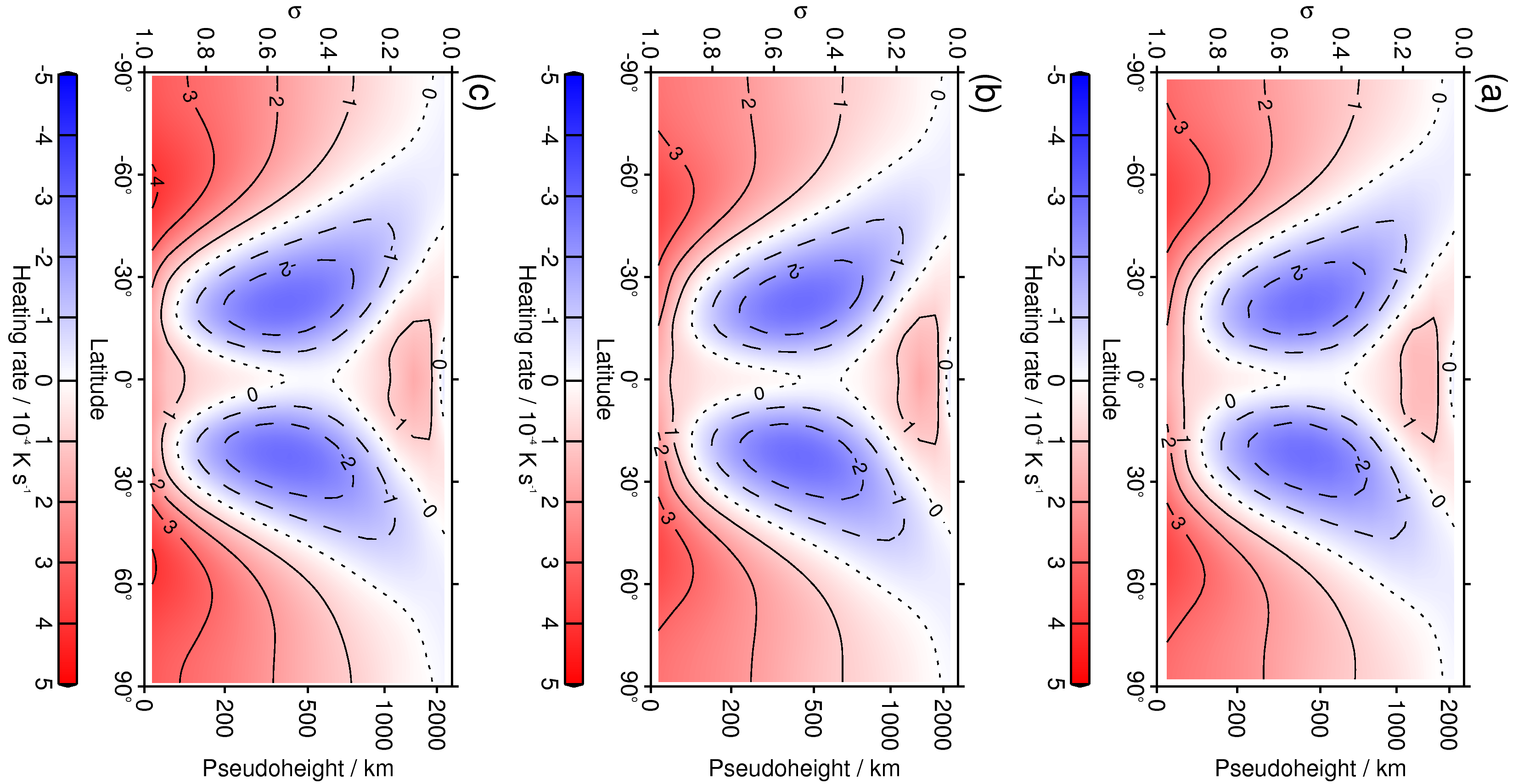}
\caption{Time-averaged zonal mean heating rate plots at different resolutions. Note the units of $10^{-4}$ K s$^{-1}$. (a) T42L15, (b) T63L20, (c) T85L20.}
\label{FigZMHeating}
\end{figure}

By subtracting the time mean temperature field from the known forcing state and dividing through by the restoration time-scale, plots of the heating rate resulting from the forcing can also be derived, as shown in Fig. \ref{FigZMHeating}. It can be more clearly seen from this figure that average net cooling is thus experienced in the equatorially symmetric regions between $0^\circ$ and $\pm 45^\circ$ N, while net heating is experienced in regions poleward of $\pm 45^\circ$ N. Heating is seen in all locations below $\sigma = 0.9$, as well as between $\pm 30^\circ$ N above $\sigma = 0.2$. The regions of negative heating correspond to the two warm regions noted previously in Fig. \ref{FigZMTemps}. The cooling regions appear to grow broader and the heating regions stronger, particularly towards the base of the model, with increasing resolution.

\begin{figure}
\includegraphics[height=\columnwidth,angle=90]{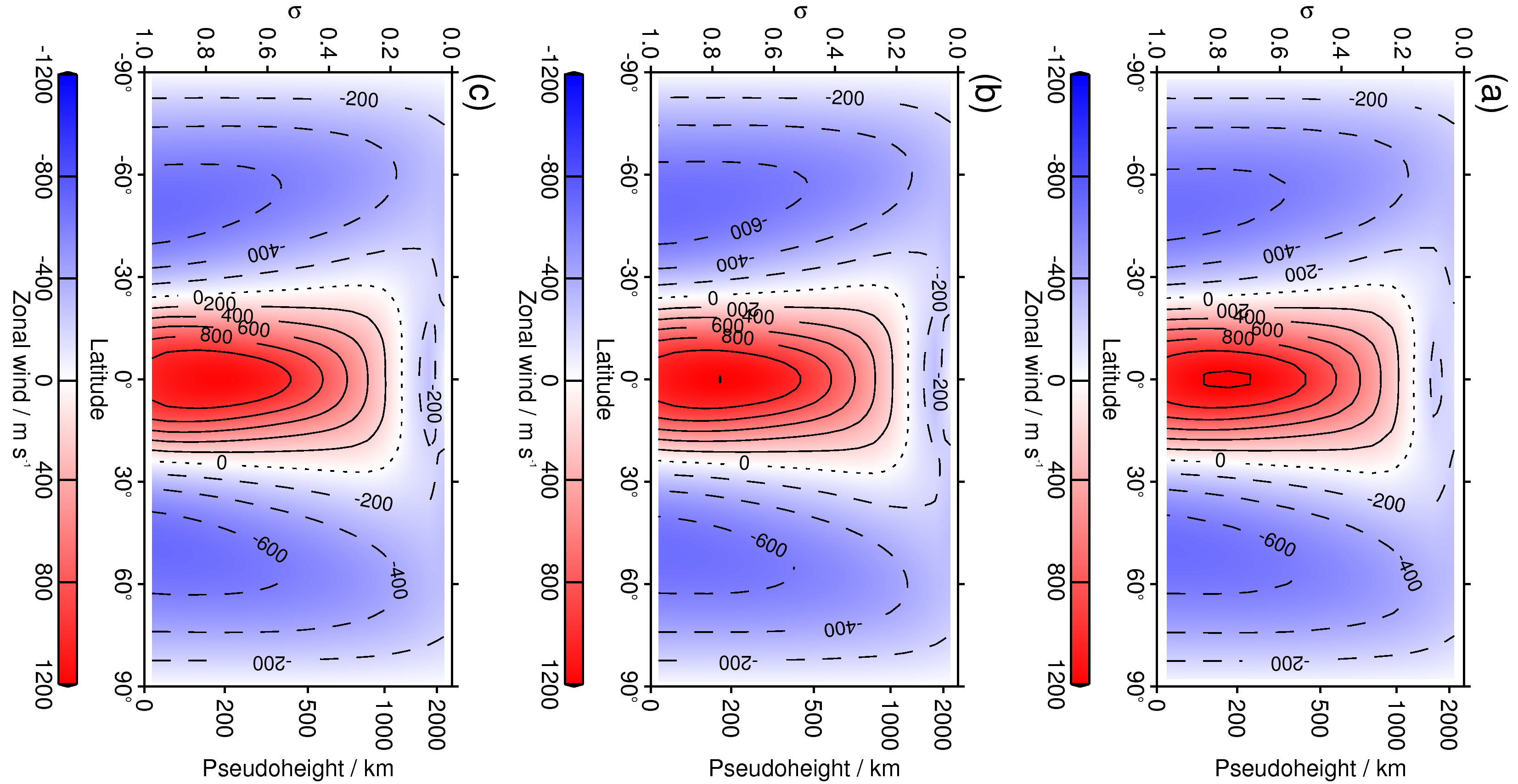}
\caption{Time average of the zonal mean zonal wind. Positive contours and colours show wind directed `out of the paper', negative contours and colours the reverse. (a) T42L15, (b) T63L20, (c) T85L20.}
\label{FigAvgZMU}
\end{figure}

The plots in Fig. \ref{FigAvgZMU} display the time-averaged, zonally averaged zonal (west-east) wind. Positive contours show wind directed west-to-east, `out of the paper', negative contours the opposite. A strongly superrotating equatorial jet is seen between $25^\circ$ S and $25^\circ$ N and below $\sigma = 0.2$, with peak windspeeds of approximately 1,200 m s$^{-1}$ in all cases, and there is a corresponding broader and weaker return flow outside this region. Maximum and minimum mean windspeeds are found to be 1220, -687 m s$^{-1}$ for the T42L15 run, 1200, -693 m s$^{-1}$ at T63L20, and 1179, -698 m s$^{-1}$ at T85L20, comparable to the \citet{intercomp2} `shallow hot Jupiter' spectral model using the same parameters. From these results, differences of order 1\% are noted when varying the resolution and hyperdiffusion. \citet{intercomp2} note that mean winds are uncertain at the level of approximately 10\% in their `deep hot Jupiter' simulation when changing parameters. The jet is unbounded at the frictionless base of the model, but closes towards the top at around $\sigma = 0.2$, with weak reverse flow above. This pattern corresponds well to that seen in both \citet{MenRauIntercomp} and the \citet{intercomp2} paper for the `shallow hot Jupiter' test case, although again, \citet{MenRauIntercomp} display only snapshots. The corresponding zonal wind snapshot for the T42L15 run is shown in Fig. \ref{FigZWsnap}, and is broadly similar to that of \citet{MenRauIntercomp} fig. 4, with the most notable difference being that the greatest zonally-averaged westward windspeed is approximately 830 as opposed to 730 m s$^{-1}$ at this particular instant in time.

\begin{figure}
\includegraphics[height=\columnwidth,angle=90]{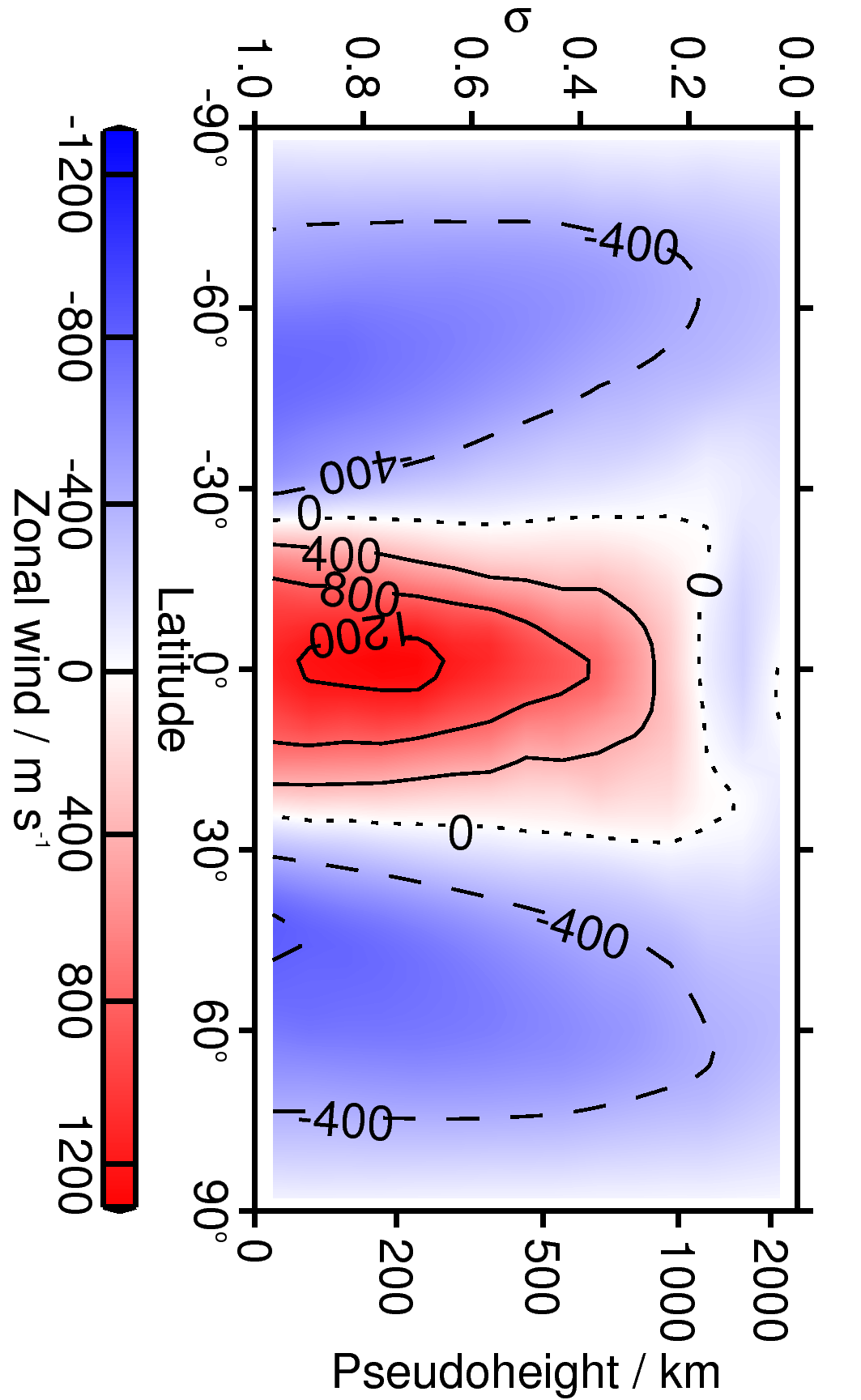}
\caption{T42L15 zonal wind `snapshot' plot, for comparison to fig. 4 of \citet{MenRauIntercomp}. Positive colours and contours show wind directed `out of the paper', negative colours and contours the reverse.}
\label{FigZWsnap}
\end{figure}

\begin{figure}
\includegraphics[height=\columnwidth,angle=90]{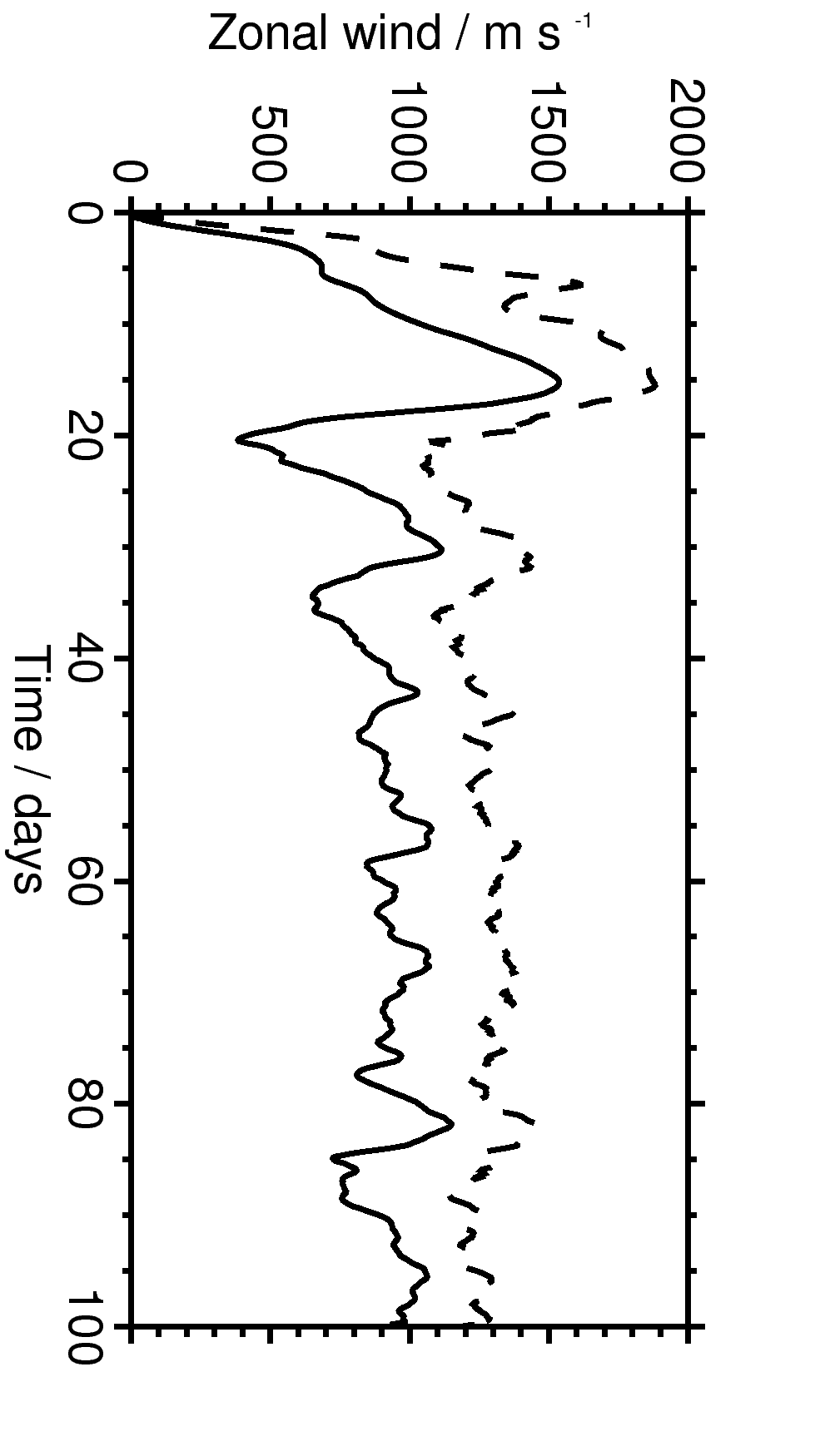}
\caption{Representative U plots for the T42L15 run, as in fig. 5 of \citet{MenRauIntercomp}. Solid line represents zonal mean wind at the equator, dashed line the equatorial maximum.}
\label{FigRepUPlots}
\end{figure}

\begin{figure}
\includegraphics[height=\columnwidth,angle=90]{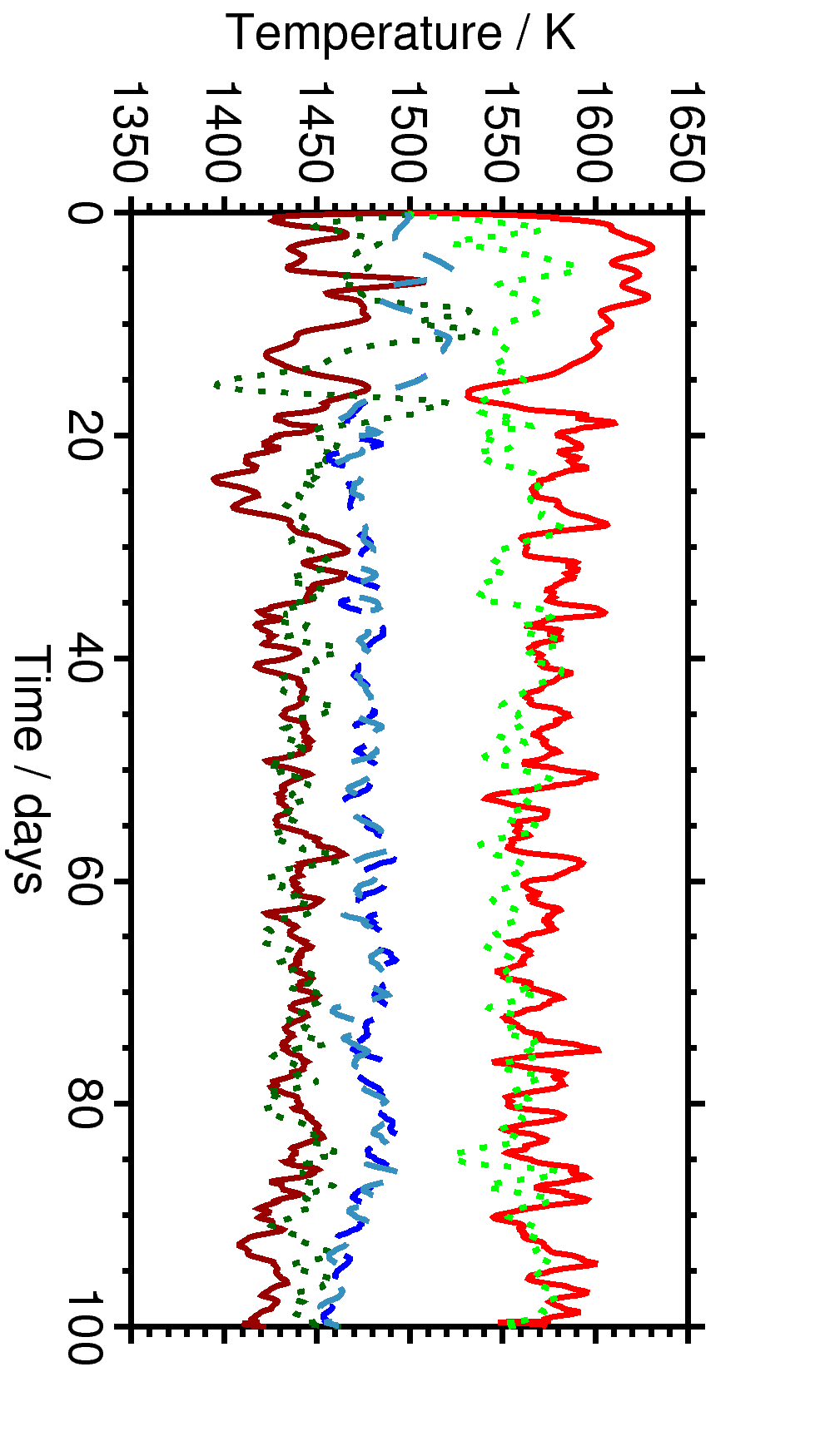}
\caption{Representative T plots for the T42L15 run, as in fig. 5 of \citet{MenRauIntercomp}. Light and dark red lines represent temperatures at the sub- and anti-stellar points, respectively, green and dark green dotted lines temperatures at the east and west limbs respectively, and blue and dark blue dashed lines temperatures at the north and south poles, respectively.}
\label{FigRepTPlots}
\end{figure}

The time evolution of the temperatures and wind speeds at various representative areas of the T42L15 run up to day 100, for comparison to fig. 5 of \citet{MenRauIntercomp}, is shown in Fig. \ref{FigRepUPlots} and Fig. \ref{FigRepTPlots}. Rather than use single points, representative areas were chosen, averaging around the longitude circles at $\pm 87.9^\circ$ N for polar values, and over the latitude band between $\pm 4.2^\circ$ N for equatorial values, which were also averaged over the longitude points between $\pm 2.8^\circ$ of the specified location (e.g. the substellar point). This enables greater reproducibility, and also helps to avoid model-specific issues in choosing the points: \textsc{puma}, for example, has neither points at $\pm 90^\circ$ N nor exactly $0^\circ$ N. For all runs, the data were sampled ten times per day. The plots shown demonstrated extremely high variability over small time-scales, and were smoothed using a simple boxcar smooth of width 1 day, or 11 records, to better display overall trends. Such high variability shows a requirement for high-frequency sampling in the creation of these plots, as daily sampling does not capture the full extent of the variability and can miss high-frequency features altogether, or produce spurious signals through aliasing.

\begin{figure*}
\begin{minipage}{165mm}
\includegraphics[height=160mm,angle=90]{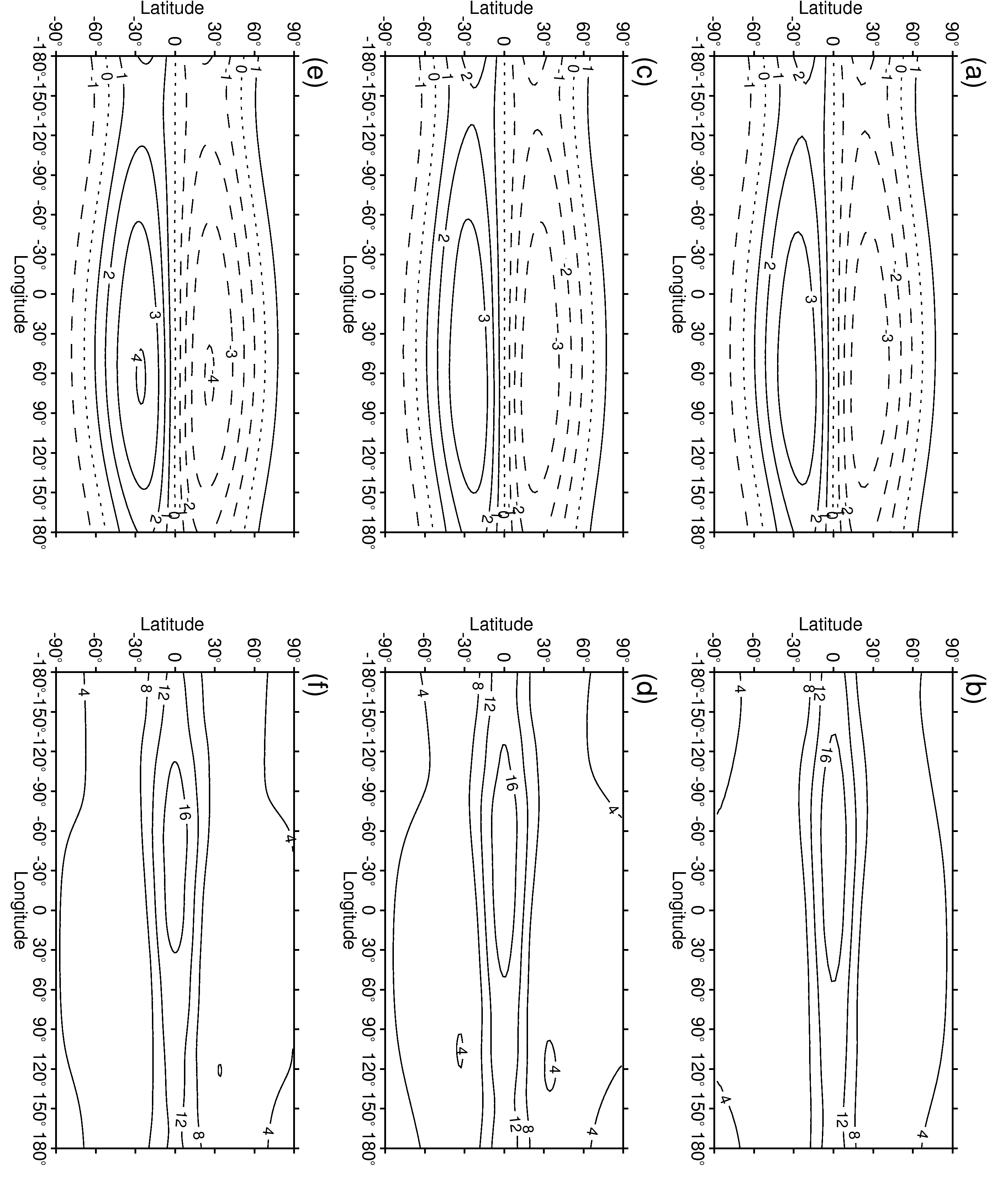}
\caption{Streamfunction mean (left) and standard deviation (right) for different resolution runs. Plots (a) and (b): T42L15 at $\sigma = 0.7$; (c) and (d): T63L20 at $\sigma = 0.725$; (e) and (f): T85L20 at $\sigma = 0.725$. The substellar point is at the centre of the image. Mean contours: $10^{10} \rmn{m}^2\rmn{s}^{-1}$; standard deviation contours: $10^9 \rmn{m}^2\rmn{s}^{-1}$}
\label{FigLLStfMeanSD}
\end{minipage}
\end{figure*}

Fig. \ref{FigLLStfMeanSD} (a), (c), and (e) display the streamfunction on the model level closest to $\sigma = 0.7$. This demonstrates the circulation of the flow, and its direction, with a large, positive streamfunction value indicating strong clockwise circulation. Its associated variance can be seen in Fig. \ref{FigLLStfMeanSD} (b), (d), and (f). It can be seen that there are two major circulation features, one to either side of the equator, centred around $60^\circ$ E of the substellar point, while the greatest variability is found on the equator around $60^\circ$ W of the substellar point. The circulation at this level strengthens slightly with increasing resolution, with its maximum value just exceeding $4 \times 10^{10}$ m$^2$ s$^{-1}$ at T85.

\begin{figure*}
\begin{minipage}{165mm}
\includegraphics[height=160mm,angle=90]{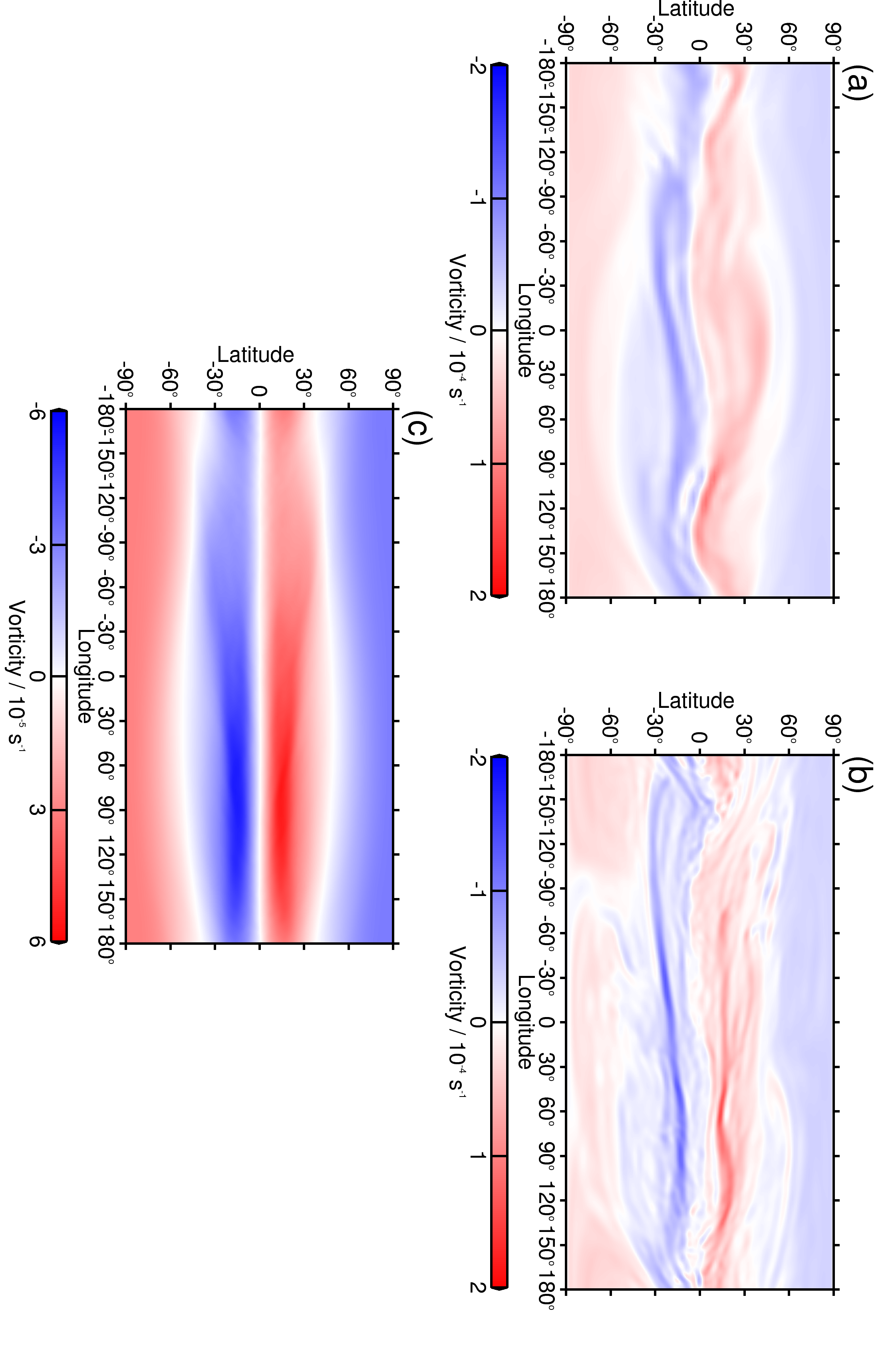}
\caption{Vorticity day-100 snapshots and time mean for T42L15 and T85L20 runs. (a): T42L15 snapshot at $\sigma = 0.7$; (b): T85L20 snapshot at $\sigma = 0.725$; (c): T85L20 time mean at $\sigma = 0.725$. The substellar point is at the centre of the image. Note the difference in scale of plot (c), at $10^{-5}$ s$^{-1}$ rather than the $10^{-4}$ s$^{-1}$ of plots (a) and (b)}
\label{FigVorticity}
\end{minipage}
\end{figure*}

Fig. \ref{FigVorticity} shows snapshots of vorticity, a measure of the local rotation at each point, on the level nearest $\sigma = 0.7$ for both T42L15 and T85L20 runs, demonstrating the impact of increasing resolution. Although a degree of structure is visible at T42, much finer, smaller-scale structure can be seen in the T85 plot, with long `streamers' of high-magnitude vorticity visible that are washed out at T42 resolution. A single time mean is shown as the time mean vorticity plots are almost identical between resolutions. Though fields such as temperature and wind appear quite similar to one another even in snapshot forms, the vorticity clearly shows the benefit of higher-resolution runs, enabling much finer detail to be resolved.

\begin{figure}
\includegraphics[height=\columnwidth,angle=90]{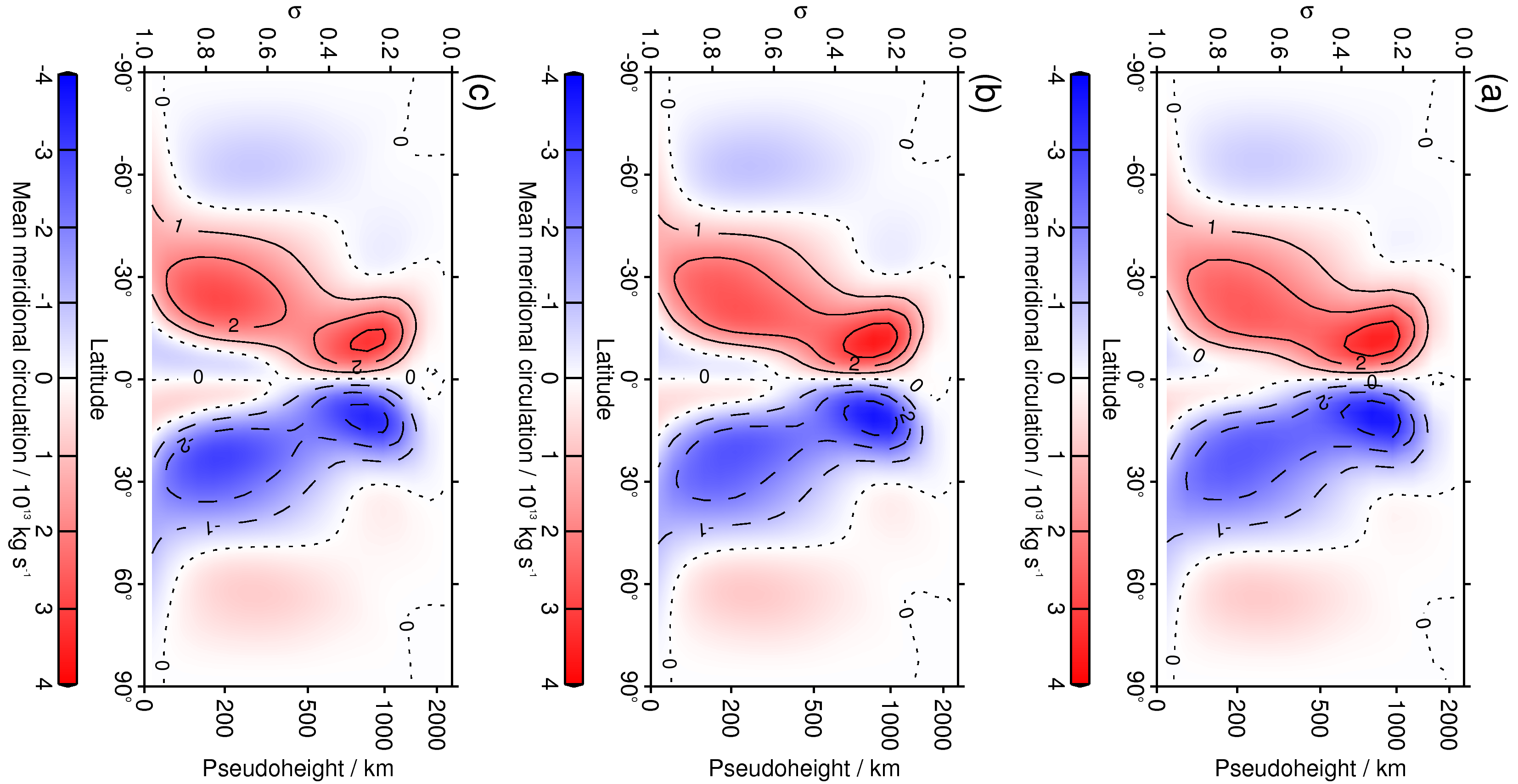}
\caption{Mean meridional circulation. Positive contours correspond to circulation in a clockwise sense, negative contours to anticlockwise circulation. (a) T42L15, (b) T63L20, (c) T85L20}
\label{FigMMCs}
\end{figure}

The plot in Fig. \ref{FigMMCs} displays the mean meridional circulation, a measure of the mass of air circulating about a given point. Positive contours indicate clockwise circulation, negative anticlockwise; a typical MMC plot for the Earth would show positive contours between $0^\circ$N and $30^\circ$N, illustrating clockwise circulation in which air rises over the equator and descends further north, followed by the inverse between $30^\circ$N and $60^\circ$N, and a further clockwise circulation between there and the pole \citep{Earthref}. Similarly, the inverse pattern is seen in the southern hemisphere. Here it can be seen that there are two main circulation features, with air descending rather than rising over the equator \citep[as expected from][]{ShowmanSuperrot}, and rising between $\pm 30^\circ$N and $\pm 60^\circ$N, with a weak reverse circulation towards the poles. The equatorial circulation contracts and weakens slightly with increasing resolution, with the two distinct peaks at different altitudes becoming more obvious

\begin{figure}
\includegraphics[height=\columnwidth,angle=90]{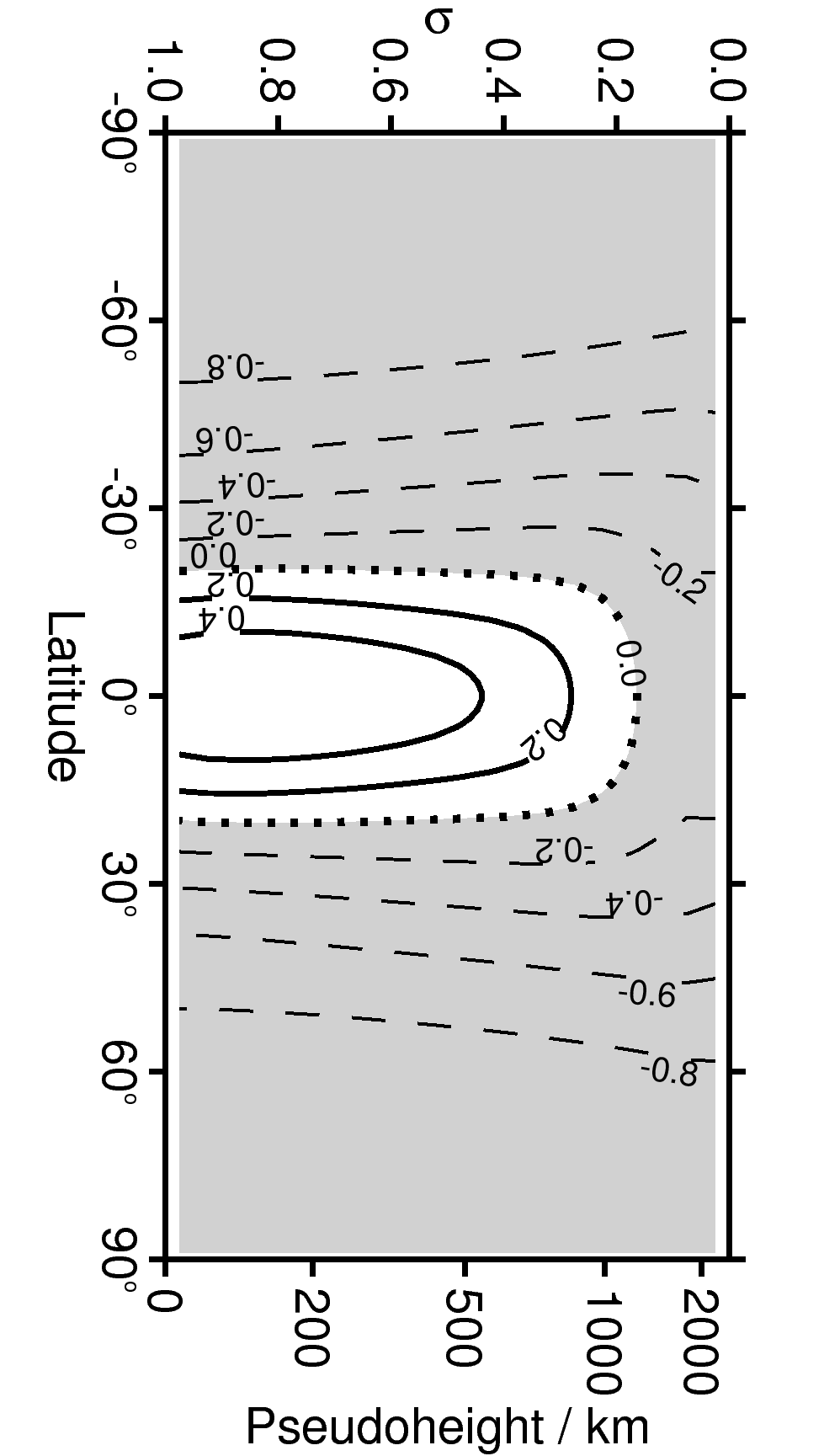}
\caption{Time-averaged local superrotation index for the T85L20 run.}
\label{FigSupRot}
\end{figure}

Fig. \ref{FigSupRot} shows the local superrotation index, which is a measure of the degree to which the angular momentum of each element of the atmosphere exceeds that which it would have in solid-body rotation. The local superrotation index $s$ is defined by
\begin{equation}
s = \overline{m} / ( \Omega a^2 ) - 1
\label{EqSindex}
\end{equation}
where $\overline{m}$ is the axial angular momentum per unit mass of the atmosphere derived from the zonal mean zonal wind $\overline{u}$, the longitudinal average of the zonal (east-west) component $u$ of the wind field  \citep{SRLsuperrot}. In general, the axial angular momentum per unit mass is given by
\begin{equation}
m = \Omega a^2 \cos^2(\phi) + u a \cos(\phi)
\label{EqAAM}
\end{equation}
A global superrotation index $S$ can also be calculated by integrating over the whole atmosphere:
\begin{equation}
S = \left( \int\int\int ( m a^2 \cos(\phi) / g ) d\lambda d\phi dp \right) / M_\rmn{0} - 1
\label{EqGobalSuperrot}
\end{equation}
where $M_\rmn{0}$ is the same integral for the atmosphere at rest, or $u = 0$.

A westerly wind (blowing west-to-east) over the equator cannot be created from an atmosphere initially at rest simply by moving air parcels from other regions of the atmosphere, since the maximum angular momentum available is that at the equator. The existence of superrotation is thus a signature of eddy processes occurring in the atmosphere, transporting angular momentum equatorward. A detailed study of superrotation under hot Jupiter conditions can be found in \citet{ShowmanSuperrot}. Only the T85L20 run is displayed, as the results for each run are visually identical. The maximum value lies between $0.56$ and $0.57$ in each case, while the minimum is $-1$ at the poles.

Detailed study of the angular momentum budget over the course of each run reveals that the global superrotation index begins at $S=0$, as expected from the model's initialisation state of zero wind. It then climbs over the first 15 days to a value of $0.033 \pm 0.003$ in each simulation, indicating that angular momentum is not fully conserved, and additional energy has been imparted to the atmosphere. With no diurnal tides, surface friction, or topographical features to provide this extra momentum, it is likely to have been acquired through model dissipation.

\begin{figure}
\includegraphics[height=\columnwidth,angle=90]{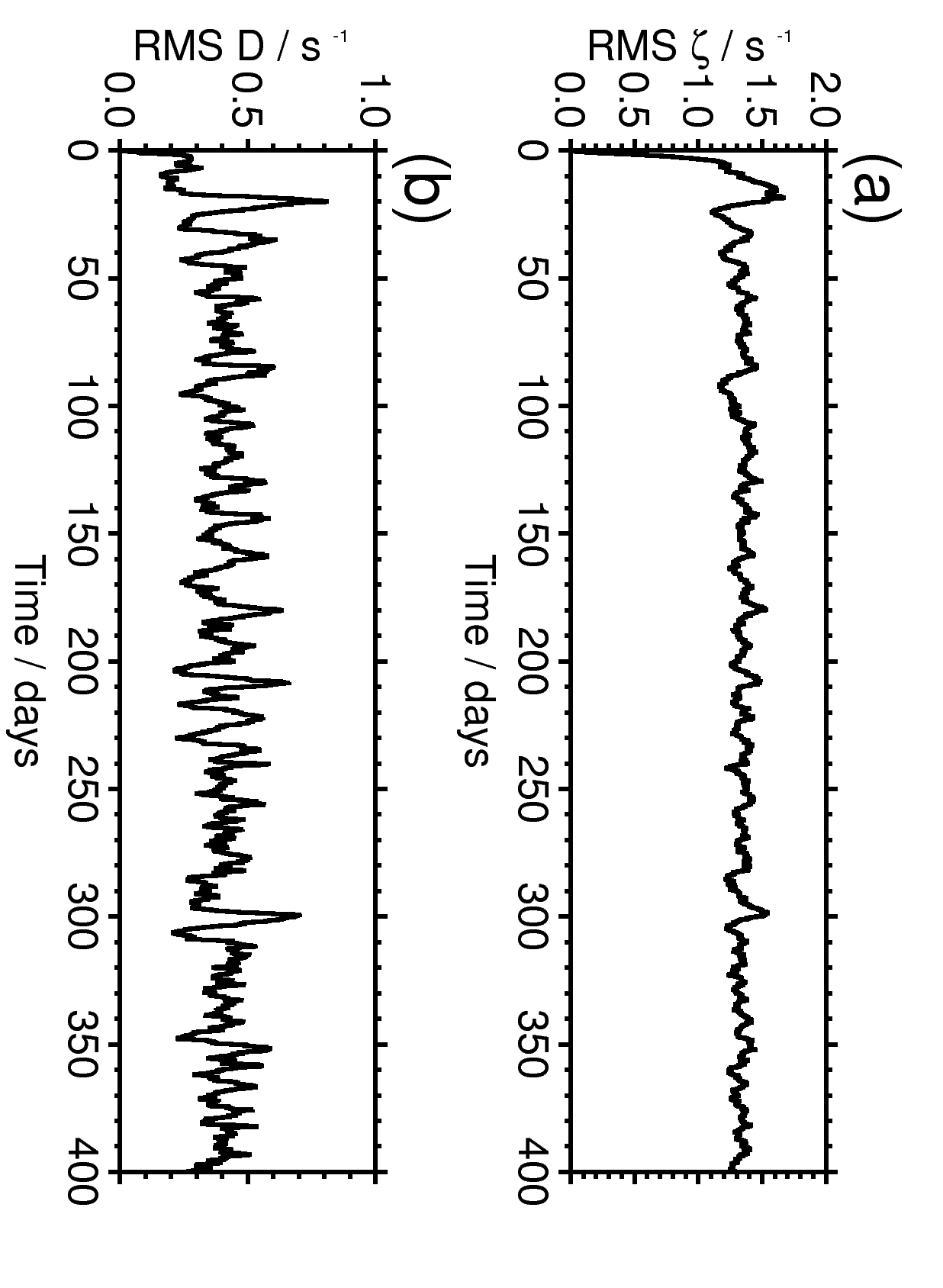}
\caption{Global statistics for T42L15. (a) Global RMS vorticity, (b) global RMS divergence.}
\label{FigT42GlobalStats}
\end{figure}

Fig. \ref{FigT42GlobalStats} shows the time evolution of the global statistics RMS vorticity and RMS divergence, which are output at each model timestep, over the initial 400-day period of the T42L15 run. These values are the root mean square of the entire global vorticity and divergence fields. A stable state is reached after 25-30 days, after which the observed degree of variability is unchanged; the same is true at the other resolutions, although the mean values reached differ. As with the representative area plots of Fig. \ref{FigRepUPlots} and Fig. \ref{FigRepTPlots}, this demonstrates that the spin-up time for \textsc{puma} under these conditions is approximately 25-30 days, after which data may be taken without fear of compromising the results with spurious spin-up effects.

\begin{figure}
\includegraphics[height=\columnwidth,angle=90]{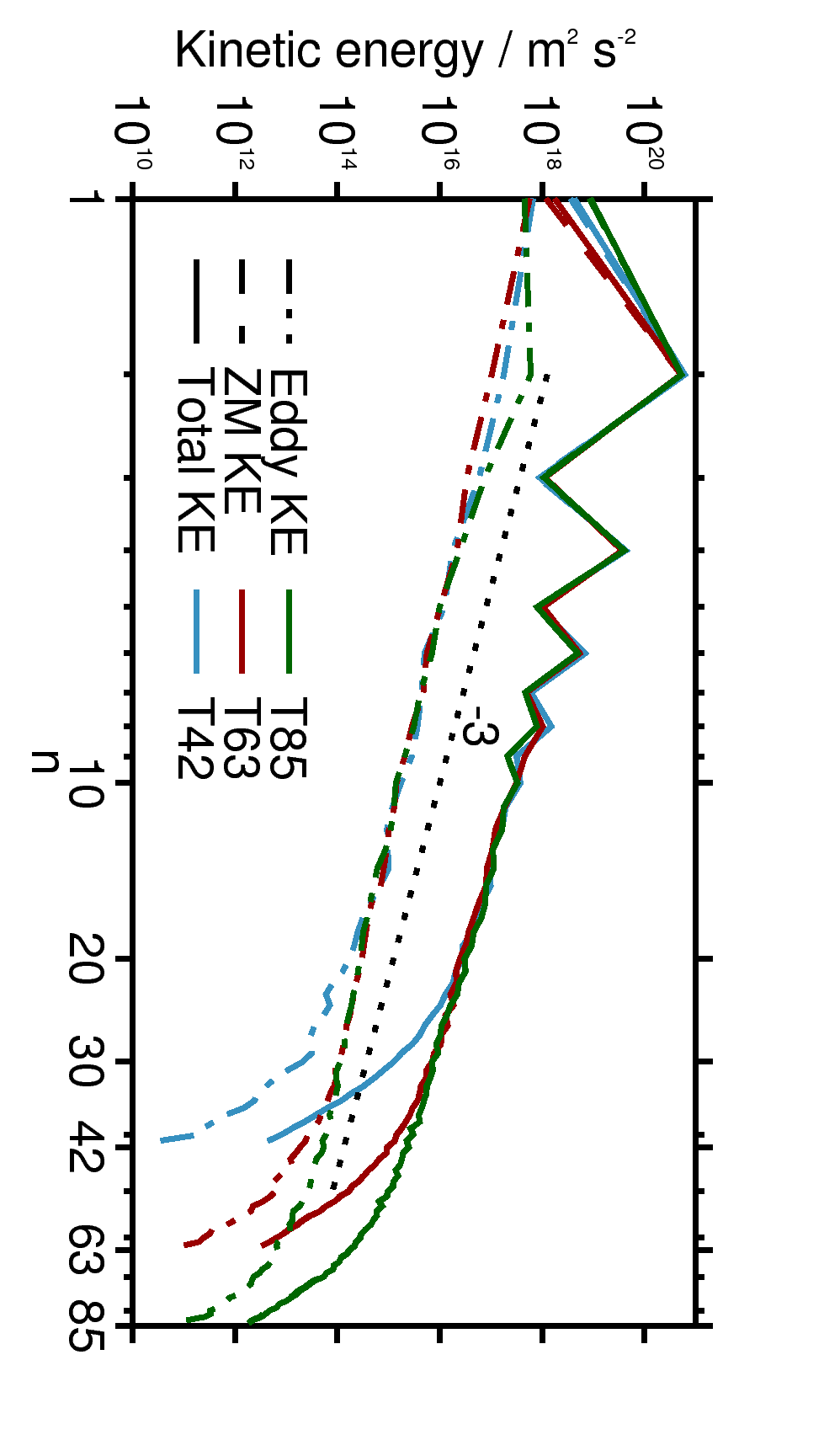}
\caption{Kinetic energy spectra for all runs. The spectra have been averaged over all levels and over a period of ten days. The blue lines denote the spectrum of the T42L15 run, the red lines T63L20, and the green lines T85L20. The total energy spectrum is shown as a solid line, while the zonal component is dashed and the eddy dash-dotted. The dotted line is a reference line with a slope of $-3$.}
\label{FigEnSpec}
\end{figure}

Fig. \ref{FigEnSpec} shows the kinetic energy spectra for three resolutions. Notably, low (large-scale), even (symmetric) wavenumbers have much higher amplitudes than their odd-valued counterparts, due to the highly equatorially symmetric, large-scale nature of the thermal forcing and final state. The dotted line has a slope of $-3$, that expected from an enstrophy-cascading range in two-dimensional turbulence \citep{minusthree}. The majority of the spectrum lies closely parallel to this line, demonstrating that this regime holds over most modelled scales. This quasi-two-dimensional regime is to be expected from the scales reachable by these studies, as the smallest resolved scale even at maximum resolution, T85, is still on the order of $10^3$ km, with flow on this scale strongly constrained by the effects of planetary rotation and atmospheric depth, rather than fully three-dimaensional turbulence \citep{HoughtonBook}. Higher wavenumbers correspond to smaller scale features, and the greater kinetic energy present at higher wavenumbers in the higher resolution runs thus results in the greater detail and higher extrema seen most clearly in the vorticity plots of Fig. \ref{FigVorticity}. In each case, the spectrum tails off sharply towards the run's wavenumber cutoff, with a slope of around $-15$. This sharp decrease near the cutoff is not linked to physical expectations and is a result of the model diffusion. While diffusive processes do naturally occur, the limitations of modelling require that they must be represented at progressively larger scales (lower wavenumbers) as the model resolution decreases, to avoid an unphysical build-up of energy at the smallest resolved scales. In a true system, this energy would continue to cascade down to ever smaller scales and eventually be dissipated; the model system, however, cannot reach such small scales, and since it conserves energy efficiently, must have additional dissipation applied.

\section{Discussion}\label{SecDiscuss}

As the results presented have shown, although `snapshot' plots are useful for capturing a view of how the simulated atmosphere is behaving, they are effectively irreproducible (contrast, for example, Fig. \ref{FigMRIEcylind} of this paper and fig. 3 of \citet{MenRauIntercomp}), and can only be compared qualitatively, rather than quantitatively. They may even be misleading, if the `snapshot' time is chosen while the model is far from the mean state. Time mean plots, as produced by \citet{intercomp2}, provide directly comparable results, as the mean flow would be expected to be very similar between any two simulations reaching the same stable state. However, the variability of the model is then lost without the addition of variance or standard deviation information. Such small-scale variability is another important point of comparison, as it demonstrates potential planetary `weather' and also shows the mean transport of such quantities as heat and momentum by transient waves. Taken together, therefore, plots of the mean and associated standard deviation are found to be more useful for the purposes of model intercomparison work.

Resolution is found to make little difference to many of the mean results studied, with similar mean temperatures and winds recorded in all cases. The most visible differences between runs are seen in the standard deviation contours, denoting different levels of variability in different regions. The persistence of the large-scale features between runs and models \citep{MenRauIntercomp, intercomp2} suggests that such features, with a powerful superrotating equatorial jet and correspondingly offset temperature maxima and minima, are likely to be observed on true extrasolar planets subject to such extreme forcing conditions. The observations available appear to support this conclusion, evidenced by the results of \citet{TempMap}, who fit a simple model of planetary brightness temperature to the phase variation of the HD 189733 light curve to obtain a temperature map of HD 189733b, \citet*{TempMap2}, who use two different methods to gain a temperature map of HD 189733b from its secondary eclipse, and \citet{WindMeas}, who detected a 2 km s$^{-1}$ CO blueshift implying strong winds flowing from the dayside to the nightside of HD 209458b. However, very few such observations have been made, and sufficient resolution to confirm or deny the presence of features at scales much below the global is still unattained. For the present, model results thus remain the only available avenue of study for the majority of features.

The analysis of Fig. \ref{FigT42GlobalStats} as well as Fig. \ref{FigRepUPlots} and Fig. \ref{FigRepTPlots} shows that the approximate spin-up time for \textsc{puma} is around 25-30 days, after which a stable state is reached and the starting conditions are effectively erased. This determines the number of records it is necessary to discard before averaging the data to avoid introducing spurious information, and may vary from model to model. This is a relatively short period of time compared to typical Earth spinup times, and extremely short in comparison to the times required to spin up a model of Jupiter itself, which receives very little external heating.

Sampling frequency can visibly alter the appearance of the representative value plots, with the daily sampling used by \citet{MenRauIntercomp} hiding the high variability seen on smaller time-scales, and use of a time filter together with a high sampling frequency is recommended.

The kinetic energy spectra at all resolutions largely follow the $-3$ law of \citet{minusthree}, but demonstrate a sharp fall in energy at scales approaching the truncation wavenumber, with a slope of $-15$ or steeper. Each increase in resolution studied produces an extension of the enstrophy-cascading range before the point at which dissipation takes over is reached. \citet{ChoForcDiss} suggest that model calculations with these large forcing amplitudes and short restoration and dissipation time-scales are likely to become over- or under-damped, and the use of energy spectra and fields such as vorticity are recommended to determine whether this is the case.

\section{Conclusions}\label{SecConclusions}

Benchmark tests of GCMs for hot Jupiters have been considered, and a variety of diagnostics produced and analysed.

`Snapshot' plots are of limited use for the purposes of an intercomparison study because the precise phase of waves depends sensitively on the initial conditions and the evolution of the flow. The use of mean and standard deviation plots is therefore encouraged. The addition of plots such as streamfunction, mean meridional circulation, and superrotation index to the simple temperature and wind fields are also suggested, as these alone do not capture all aspects of the simulation, although more care is required in interpreting some such plots.

The use of high sampling frequencies is also found to be important, with the degree of variability seen in the `representative area' plots of Fig. \ref{FigRepUPlots} and Fig. \ref{FigRepTPlots} demonstrably dependent on the frequency at which the data is sampled. A sampling frequency of at least ten records per day is suggested, and the data time-filtered.

The resolution of the model runs is found to have little effect on the overall mean and variability of the data. However, much finer detail is captured at high resolution, as illustrated in the vorticity plots of Fig. \ref{FigVorticity}. Smoother fields such as temperature show less visible effect of this increased resolution. The energy spectra show that the higher-resolution models continue to follow the $-3$ law to smaller scales before dissipation takes effect, demonstrating that worthwhile data can be gained by increasing the resolution.

The large-scale features of the temperature and wind fields correspond in form and magnitude to those of \citet{MenRauIntercomp} and \citet{intercomp2}, to the extent that those can be determined from the results provided. This reinforces the growing consensus that such features, with a `hotspot' offset from the substellar point by a strongly superrotating equatorial jet, are likely to be found on a variety of hot Jupiter type exoplanets.

\section{Acknowledgements}

The authors are grateful for the support of STFC. VLB additionally thanks both STFC and the Open University's Charter Studentships scheme, and would also like to thank Yixiong Wang at Oxford University, whose correspondence has been helpful. We also thank the anonymous referee, whose detailed comments helped to improve various aspects of this paper.

\end{document}